\RequirePackage{lineno}

\documentclass[aps,prl,twocolumn,superscriptaddress,showpacs]{revtex4-1}

\usepackage{amssymb,graphicx,epsfig}
\usepackage{latexsym,amsmath,amssymb,mathrsfs,multirow,subfigure}
\usepackage[english]{babel}
\usepackage{rotating,graphics,overpic,textcomp,xspace,lineno}
\usepackage[urlcolor=blue, 
            colorlinks=true,
            linkcolor=blue,  
            bookmarks,       
            bookmarksnumbered=true]{hyperref}

\newcommand{\afbtt}{A_{\rm{FB}}^{t\bar{t}}}

\newcommand{\Me}{\mbox{$E\kern-0.50em\raise0.10ex\hbox{/}$}}

\newcommand{\Mtt}{m_{\ttbar}}
\newcommand{\Dy}{\Delta y_{\ttbar}}

\newcommand{\MetVec}{\mbox{$\vec E\kern-0.50em\raise0.10ex\hbox{/}_{T}$}}
\newcommand{\MetVecRaw}{\mbox{$\vec E\kern-0.50em\raise0.10ex\hbox{/}^{raw}_{T}$}}
\newcommand{\Met}{\mbox{$E\kern-0.50em\raise0.10ex\hbox{/}_{T}$}}
\newcommand{\Mex}{\mbox{$E\kern-0.50em\raise0.10ex\hbox{/}_{x}$}}
\newcommand{\Mey}{\mbox{$E\kern-0.50em\raise0.10ex\hbox{/}_{y}$}}
\newcommand{\Mexy}{\mbox{$E\kern-0.50em\raise0.10ex\hbox{/}_{x,y}$}}
\newcommand{\MeRaw}{\mbox{$E\kern-0.50em\raise0.10ex\hbox{/}_{T}^{raw}$}}
\newcommand{\MetSpec}{\mbox{$E\kern-0.50em\raise0.10ex\hbox{/}_{T}^{spec}$}}

\newcommand{\MetSig}{\mbox{$E\kern-0.50em\raise0.10ex\hbox{/}_{T}^{sig}$}}

\newcommand{\ttbar}{t \bar{t}}

\newcommand{\gevcc}{{\rm GeV}/c^2}

\newcommand{\MET}{\mbox{$E\kern-0.50em\raise0.10ex\hbox{/}_{T}$}}
\newcommand{\met}{\mbox{$E\kern-0.50em\raise0.10ex\hbox{/}_{T}$}}
\newcommand{\METzero}{\mbox{$E\kern-0.50em\raise0.10ex\hbox{/}_{T}^0$}}
\newcommand{\vecMET}{\mbox{$\vec{E}\kern-0.50em\raise0.10ex\hbox{/}_{T}$}}
\newcommand{\METSPEC}{\mbox{$E\kern-0.50em\raise0.10ex\hbox{/}_{Tspec}$}}

\newcommand{\ppbar}{p \overline{p}}

\def\slantfrac#1#2{\kern.1em^{#1}\kern-.3em/\kern-.1em_{#2}}

\newcommand{\deta}{\Delta\eta}
\newcommand{\afblep}{A_{\rm{FB}}^{\ell}}

\newcommand{\afbdeta}{A_{\rm{FB}}^{\ell\ell}}

\newcommand\T{\rule{0pt}{2.6ex}}       

\mathchardef\mhyphen="2D

\begin{document}

\onecolumngrid
\hfill \mbox{FERMILAB-PUB-17-379-E}  
\twocolumngrid

\title{Combined Forward-Backward Asymmetry Measurements in Top-Antitop Quark Production at the Tevatron}

\affiliation{Institute of Physics, Academia Sinica, Taipei, Taiwan 11529, Republic of China}
\affiliation{Argonne National Laboratory, Argonne, Illinois 60439, USA}
\affiliation{University of Athens, 157 71 Athens, Greece}
\affiliation{Institut de Fisica d'Altes Energies, ICREA, Universitat Autonoma de Barcelona, E-08193, Bellaterra (Barcelona), Spain}
\affiliation{Baylor University, Waco, Texas 76798, USA}
\affiliation{Istituto Nazionale di Fisica Nucleare Bologna, \ensuremath{^{aaa}}University of Bologna, I-40127 Bologna, Italy}
\affiliation{University of California, Davis, Davis, California 95616, USA}
\affiliation{University of California, Los Angeles, Los Angeles, California 90024, USA}
\affiliation{Instituto de Fisica de Cantabria, CSIC-University of Cantabria, 39005 Santander, Spain}
\affiliation{Carnegie Mellon University, Pittsburgh, Pennsylvania 15213, USA}
\affiliation{Enrico Fermi Institute, University of Chicago, Chicago, Illinois 60637, USA}
\affiliation{Comenius University, 842 48 Bratislava, Slovakia; Institute of Experimental Physics, 040 01 Kosice, Slovakia}
\affiliation{Joint Institute for Nuclear Research, RU-141980 Dubna, Russia}
\affiliation{Duke University, Durham, North Carolina 27708, USA}
\affiliation{Fermi National Accelerator Laboratory, Batavia, Illinois 60510, USA}
\affiliation{University of Florida, Gainesville, Florida 32611, USA}
\affiliation{Laboratori Nazionali di Frascati, Istituto Nazionale di Fisica Nucleare, I-00044 Frascati, Italy}
\affiliation{University of Geneva, CH-1211 Geneva 4, Switzerland}
\affiliation{Glasgow University, Glasgow G12 8QQ, United Kingdom}
\affiliation{Harvard University, Cambridge, Massachusetts 02138, USA}
\affiliation{Division of High Energy Physics, Department of Physics, University of Helsinki, FIN-00014, Helsinki, Finland; Helsinki Institute of Physics, FIN-00014, Helsinki, Finland}
\affiliation{University of Illinois, Urbana, Illinois 61801, USA}
\affiliation{The Johns Hopkins University, Baltimore, Maryland 21218, USA}
\affiliation{Institut f\"{u}r Experimentelle Kernphysik, Karlsruhe Institute of Technology, D-76131 Karlsruhe, Germany}
\affiliation{Center for High Energy Physics: Kyungpook National University, Daegu 702-701, Korea; Seoul National University, Seoul 151-742, Korea; Sungkyunkwan University, Suwon 440-746, Korea; Korea Institute of Science and Technology Information, Daejeon 305-806, Korea; Chonnam National University, Gwangju 500-757, Korea; Chonbuk National University, Jeonju 561-756, Korea; Ewha Womans University, Seoul, 120-750, Korea}
\affiliation{Ernest Orlando Lawrence Berkeley National Laboratory, Berkeley, California 94720, USA}
\affiliation{University of Liverpool, Liverpool L69 7ZE, United Kingdom}
\affiliation{University College London, London WC1E 6BT, United Kingdom}
\affiliation{Centro de Investigaciones Energeticas Medioambientales y Tecnologicas, E-28040 Madrid, Spain}
\affiliation{Massachusetts Institute of Technology, Cambridge, Massachusetts 02139, USA}
\affiliation{University of Michigan, Ann Arbor, Michigan 48109, USA}
\affiliation{Michigan State University, East Lansing, Michigan 48824, USA}
\affiliation{Institute for Theoretical and Experimental Physics, ITEP, Moscow 117259, Russia}
\affiliation{University of New Mexico, Albuquerque, New Mexico 87131, USA}
\affiliation{The Ohio State University, Columbus, Ohio 43210, USA}
\affiliation{Okayama University, Okayama 700-8530, Japan}
\affiliation{Osaka City University, Osaka 558-8585, Japan}
\affiliation{University of Oxford, Oxford OX1 3RH, United Kingdom}
\affiliation{Istituto Nazionale di Fisica Nucleare, Sezione di Padova, \ensuremath{^{bbb}}University of Padova, I-35131 Padova, Italy}
\affiliation{University of Pennsylvania, Philadelphia, Pennsylvania 19104, USA}
\affiliation{Istituto Nazionale di Fisica Nucleare Pisa, \ensuremath{^{ccc}}University of Pisa, \ensuremath{^{ddd}}University of Siena, \ensuremath{^{eee}}Scuola Normale Superiore, I-56127 Pisa, Italy, \ensuremath{^{fff}}INFN Pavia, I-27100 Pavia, Italy, \ensuremath{^{ggg}}University of Pavia, I-27100 Pavia, Italy}
\affiliation{University of Pittsburgh, Pittsburgh, Pennsylvania 15260, USA}
\affiliation{Purdue University, West Lafayette, Indiana 47907, USA}
\affiliation{University of Rochester, Rochester, New York 14627, USA}
\affiliation{The Rockefeller University, New York, New York 10065, USA}
\affiliation{Istituto Nazionale di Fisica Nucleare, Sezione di Roma 1, \ensuremath{^{hhh}}Sapienza Universit\`{a} di Roma, I-00185 Roma, Italy}
\affiliation{Mitchell Institute for Fundamental Physics and Astronomy, Texas A\&M University, College Station, Texas 77843, USA}
\affiliation{Istituto Nazionale di Fisica Nucleare Trieste, \ensuremath{^{iii}}Gruppo Collegato di Udine, \ensuremath{^{jjj}}University of Udine, I-33100 Udine, Italy, \ensuremath{^{kkk}}University of Trieste, I-34127 Trieste, Italy}
\affiliation{University of Tsukuba, Tsukuba, Ibaraki 305, Japan}
\affiliation{Tufts University, Medford, Massachusetts 02155, USA}
\affiliation{Waseda University, Tokyo 169, Japan}
\affiliation{Wayne State University, Detroit, Michigan 48201, USA}
\affiliation{University of Wisconsin-Madison, Madison, Wisconsin 53706, USA}
\affiliation{Yale University, New Haven, Connecticut 06520, USA}
\affiliation{LAFEX, Centro Brasileiro de Pesquisas F\'{i}sicas, Rio de Janeiro, RJ 22290, Brazil}
\affiliation{Universidade do Estado do Rio de Janeiro, Rio de Janeiro, RJ 20550, Brazil}
\affiliation{Universidade Federal do ABC, Santo Andr\'{e}, SP 09210, Brazil}
\affiliation{University of Science and Technology of China, Hefei 230026, People's Republic of China}
\affiliation{Universidad de los Andes, Bogot\'{a}, 111711, Colombia}
\affiliation{Charles University, Faculty of Mathematics and Physics, Center for Particle Physics, 116 36 Prague 1, Czech Republic}
\affiliation{Czech Technical University in Prague, 116 36 Prague 6, Czech Republic}
\affiliation{Institute of Physics, Academy of Sciences of the Czech Republic, 182 21 Prague, Czech Republic}
\affiliation{Universidad San Francisco de Quito, Quito 170157, Ecuador}
\affiliation{LPC, Universit\'{e} Blaise Pascal, CNRS/IN2P3, Clermont, F-63178 Aubi\`ere Cedex, France}
\affiliation{LPSC, Universit\'{e} Joseph Fourier Grenoble 1, CNRS/IN2P3, Institut National Polytechnique de Grenoble, F-38026 Grenoble Cedex, France}
\affiliation{CPPM, Aix-Marseille Universit\'{e}, CNRS/IN2P3, F-13288 Marseille Cedex 09, France}
\affiliation{LAL, Univ. Paris-Sud, CNRS/IN2P3, Universit\'{e} Paris-Saclay, F-91898 Orsay Cedex, France}
\affiliation{LPNHE, Universit\'{e}s Paris VI and VII, CNRS/IN2P3, F-75005 Paris, France}
\affiliation{CEA Saclay, Irfu, SPP, F-91191 Gif-Sur-Yvette Cedex, France}
\affiliation{IPHC, Universit\'{e} de Strasbourg, CNRS/IN2P3, F-67037 Strasbourg, France}
\affiliation{IPNL, Universit\'{e} Lyon 1, CNRS/IN2P3, F-69622 Villeurbanne Cedex, France and Universit\'{e} de Lyon, F-69361 Lyon CEDEX 07, France}
\affiliation{III. Physikalisches Institut A, RWTH Aachen University, 52056 Aachen, Germany}
\affiliation{Physikalisches Institut, Universit\"{a}t Freiburg, 79085 Freiburg, Germany}
\affiliation{II. Physikalisches Institut, Georg-August-Universit\"{a}t G\"{o}ttingen, 37073 G\"{o}ttingen, Germany}
\affiliation{Institut f\"{u}r Physik, Universit\"{a}t Mainz, 55099 Mainz, Germany}
\affiliation{Ludwig-Maximilians-Universit\"{a}t M\"{u}nchen, 80539 M\"{u}nchen, Germany}
\affiliation{Panjab University, Chandigarh 160014, India}
\affiliation{Delhi University, Delhi-110 007, India}
\affiliation{Tata Institute of Fundamental Research, Mumbai-400 005, India}
\affiliation{University College Dublin, Dublin 4, Ireland}
\affiliation{Korea Detector Laboratory, Korea University, Seoul, 02841, Korea}
\affiliation{CINVESTAV, Mexico City 07360, Mexico}
\affiliation{Nikhef, Science Park, 1098 XG Amsterdam, the Netherlands}
\affiliation{Radboud University Nijmegen, 6525 AJ Nijmegen, the Netherlands}
\affiliation{Joint Institute for Nuclear Research, RU-141980 Dubna, Russia}
\affiliation{Institute for Theoretical and Experimental Physics, ITEP, Moscow 117259, Russia}
\affiliation{Moscow State University, Moscow 119991, Russia}
\affiliation{Institute for High Energy Physics, Protvino, Moscow region 142281, Russia}
\affiliation{Petersburg Nuclear Physics Institute, St. Petersburg 188300, Russia}
\affiliation{Instituci\'{o} Catalana de Recerca i Estudis Avan\c{c}ats (ICREA) and Institut de F\'{i}sica d'Altes Energies (IFAE), 08193 Bellaterra (Barcelona), Spain}
\affiliation{Uppsala University, 751 05 Uppsala, Sweden}
\affiliation{Taras Shevchenko National University of Kyiv, Kiev, 01601, Ukraine}
\affiliation{Lancaster University, Lancaster LA1 4YB, United Kingdom}
\affiliation{Imperial College London, London SW7 2AZ, United Kingdom}
\affiliation{The University of Manchester, Manchester M13 9PL, United Kingdom}
\affiliation{University of Arizona, Tucson, Arizona 85721, USA}
\affiliation{University of California Riverside, Riverside, California 92521, USA}
\affiliation{Florida State University, Tallahassee, Florida 32306, USA}
\affiliation{Fermi National Accelerator Laboratory, Batavia, Illinois 60510, USA}
\affiliation{University of Illinois at Chicago, Chicago, Illinois 60607, USA}
\affiliation{Northern Illinois University, DeKalb, Illinois 60115, USA}
\affiliation{Northwestern University, Evanston, Illinois 60208, USA}
\affiliation{Indiana University, Bloomington, Indiana 47405, USA}
\affiliation{Purdue University Calumet, Hammond, Indiana 46323, USA}
\affiliation{University of Notre Dame, Notre Dame, Indiana 46556, USA}
\affiliation{Iowa State University, Ames, Iowa 50011, USA}
\affiliation{University of Kansas, Lawrence, Kansas 66045, USA}
\affiliation{Louisiana Tech University, Ruston, Louisiana 71272, USA}
\affiliation{Northeastern University, Boston, Massachusetts 02115, USA}
\affiliation{University of Michigan, Ann Arbor, Michigan 48109, USA}
\affiliation{Michigan State University, East Lansing, Michigan 48824, USA}
\affiliation{University of Mississippi, University, Mississippi 38677, USA}
\affiliation{University of Nebraska, Lincoln, Nebraska 68588, USA}
\affiliation{Rutgers University, Piscataway, New Jersey 08855, USA}
\affiliation{Princeton University, Princeton, New Jersey 08544, USA}
\affiliation{State University of New York, Buffalo, New York 14260, USA}
\affiliation{University of Rochester, Rochester, New York 14627, USA}
\affiliation{State University of New York, Stony Brook, New York 11794, USA}
\affiliation{Brookhaven National Laboratory, Upton, New York 11973, USA}
\affiliation{Langston University, Langston, Oklahoma 73050, USA}
\affiliation{University of Oklahoma, Norman, Oklahoma 73019, USA}
\affiliation{Oklahoma State University, Stillwater, Oklahoma 74078, USA}
\affiliation{Oregon State University, Corvallis, Oregon 97331, USA}
\affiliation{Brown University, Providence, Rhode Island 02912, USA}
\affiliation{University of Texas, Arlington, Texas 76019, USA}
\affiliation{Southern Methodist University, Dallas, Texas 75275, USA}
\affiliation{Rice University, Houston, Texas 77005, USA}
\affiliation{University of Virginia, Charlottesville, Virginia 22904, USA}
\affiliation{University of Washington, Seattle, Washington 98195, USA}

\author{T.~Aaltonen\ensuremath{^{\dagger}}}
\affiliation{Division of High Energy Physics, Department of Physics, University of Helsinki, FIN-00014, Helsinki, Finland; Helsinki Institute of Physics, FIN-00014, Helsinki, Finland}
\author{V.M.~Abazov \ensuremath{^{\ddagger}}}
\affiliation{Joint Institute for Nuclear Research, RU-141980 Dubna, Russia}
\author{B.~Abbott \ensuremath{^{\ddagger}}}
\affiliation{University of Oklahoma, Norman, Oklahoma 73019, USA}
\author{B.S.~Acharya \ensuremath{^{\ddagger}}}
\affiliation{Tata Institute of Fundamental Research, Mumbai-400 005, India}
\author{M.~Adams \ensuremath{^{\ddagger}}}
\affiliation{University of Illinois at Chicago, Chicago, Illinois 60607, USA}
\author{T.~Adams \ensuremath{^{\ddagger}}}
\affiliation{Florida State University, Tallahassee, Florida 32306, USA}
\author{J.P.~Agnew \ensuremath{^{\ddagger}}}
\affiliation{The University of Manchester, Manchester M13 9PL, United Kingdom}
\author{G.D.~Alexeev \ensuremath{^{\ddagger}}}
\affiliation{Joint Institute for Nuclear Research, RU-141980 Dubna, Russia}
\author{G.~Alkhazov \ensuremath{^{\ddagger}}}
\affiliation{Petersburg Nuclear Physics Institute, St. Petersburg 188300, Russia}
\author{A.~Alton \ensuremath{^{\ddagger}}\ensuremath{^{kk}}}
\affiliation{University of Michigan, Ann Arbor, Michigan 48109, USA}
% ...

\collaboration{CDF Collaboration}
\altaffiliation[With visitors from]{
\ensuremath{^{a}}University of British Columbia, Vancouver, BC V6T 1Z1, Canada,
\ensuremath{^{b}}Istituto Nazionale di Fisica Nucleare, Sezione di Cagliari, 09042 Monserrato (Cagliari), Italy,
\ensuremath{^{c}}University of California Irvine, Irvine, CA 92697, USA,
\ensuremath{^{d}}Institute of Physics, Academy of Sciences of the Czech Republic, 182 21, Czech Republic,
\ensuremath{^{e}}CERN, CH-1211 Geneva, Switzerland,
\ensuremath{^{f}}Cornell University, Ithaca, NY 14853, USA,
\ensuremath{^{g}}University of Cyprus, Nicosia CY-1678, Cyprus,
\ensuremath{^{h}}Office of Science, U.S. Department of Energy, Washington, DC 20585, USA,
\ensuremath{^{i}}University College Dublin, Dublin 4, Ireland,
\ensuremath{^{j}}ETH, 8092 Z\"{u}rich, Switzerland,
\ensuremath{^{k}}University of Fukui, Fukui City, Fukui Prefecture, Japan 910-0017,
\ensuremath{^{l}}Universidad Iberoamericana, Lomas de Santa Fe, M\'{e}xico, C.P. 01219, Distrito Federal,
\ensuremath{^{m}}University of Iowa, Iowa City, IA 52242, USA,
\ensuremath{^{n}}Kinki University, Higashi-Osaka City, Japan 577-8502,
\ensuremath{^{o}}Kansas State University, Manhattan, KS 66506, USA,
\ensuremath{^{p}}Brookhaven National Laboratory, Upton, NY 11973, USA,
\ensuremath{^{q}}Istituto Nazionale di Fisica Nucleare, Sezione di Lecce, Via Arnesano, I-73100 Lecce, Italy,
\ensuremath{^{r}}Queen Mary, University of London, London, E1 4NS, United Kingdom,
\ensuremath{^{s}}University of Melbourne, Victoria 3010, Australia,
\ensuremath{^{t}}Muons, Inc., Batavia, IL 60510, USA,
\ensuremath{^{u}}Nagasaki Institute of Applied Science, Nagasaki 851-0193, Japan,
\ensuremath{^{v}}National Research Nuclear University, Moscow 115409, Russia,
\ensuremath{^{w}}Northwestern University, Evanston, IL 60208, USA,
\ensuremath{^{x}}University of Notre Dame, Notre Dame, IN 46556, USA,
\ensuremath{^{y}}Universidad de Oviedo, E-33007 Oviedo, Spain,
\ensuremath{^{z}}CNRS-IN2P3, Paris, F-75205 France,
\ensuremath{^{aa}}Universidad Tecnica Federico Santa Maria, 110v Valparaiso, Chile,
\ensuremath{^{bb}}Sejong University, Seoul 143-747, Korea,
\ensuremath{^{cc}}The University of Jordan, Amman 11942, Jordan,
\ensuremath{^{dd}}Universite catholique de Louvain, 1348 Louvain-La-Neuve, Belgium,
\ensuremath{^{ee}}University of Z\"{u}rich, 8006 Z\"{u}rich, Switzerland,
\ensuremath{^{ff}}Massachusetts General Hospital, Boston, MA 02114 USA,
\ensuremath{^{gg}}Harvard Medical School, Boston, MA 02114 USA,
\ensuremath{^{hh}}Hampton University, Hampton, VA 23668, USA,
\ensuremath{^{ii}}Los Alamos National Laboratory, Los Alamos, NM 87544, USA,
\ensuremath{^{jj}}Universit\`{a} degli Studi di Napoli Federico II, I-80138 Napoli, Italy
}
\noaffiliation
\collaboration{D0 Collaboration}
\altaffiliation[With visitors from]{
\ensuremath{^{kk}}Augustana University, Sioux Falls, SD 57197, USA,
\ensuremath{^{ll}}The University of Liverpool, Liverpool L69 3BX, UK,
\ensuremath{^{mm}}Deutsches Elektronen-Synchrotron (DESY), Notkestrase 85, Germany,
\ensuremath{^{nn}}Consejo Nacional de Ciencia y Tecnologia (Conacyt), M-03940 Mexico City, Mexico,
\ensuremath{^{oo}}SLAC, Menlo Park, CA 94025, USA,
\ensuremath{^{pp}}University College London, London WC1E 6BT, UK,
\ensuremath{^{qq}}Centro de Investigacion en Computacion - IPN, CP 07738 Mexico City, Mexico,
\ensuremath{^{rr}}Universidade Estadual Paulista, S\~{a}o Paulo, SP 01140, Brazil,
\ensuremath{^{ss}}Karlsruher Institut f\"{u}r Technologie (KIT) - Steinbuch Centre for Computing (SCC), D-76128 Karlsruher, Germany,
\ensuremath{^{tt}}Office of Science, U.S. Department of Energy, Washington, D.C. 20585, USA,
\ensuremath{^{uu}}American Association for the Advancement of Science, Washington, D.C. 20005, USA,
\ensuremath{^{vv}}National Academy of Science of Ukraine (NASU) - Kiev Institute for Nuclear Research (KINR), Kyiv 03680, Ukraine,
\ensuremath{^{ww}}University of Maryland, College Park, MD 20742, USA,
\ensuremath{^{xx}}European Organization for Nuclear Research (CERN), CH-1211 Gen\'{e}ve 23, Switzerland,
\ensuremath{^{yy}}Purdue University, West Lafayette, IN 47907, USA,
\ensuremath{^{zz}}Institute of Physics, Belgrade, CS-11080 Belgrade, Serbia
}
\noaffiliation

\date{September 13, 2017}

\vspace*{2.0cm}

\begin{abstract}
The CDF and D0 experiments at the Fermilab Tevatron have measured the asymmetry between yields of forward- and backward-produced top and antitop quarks based on their rapidity difference and the asymmetry between their decay leptons. These measurements use the full data sets collected in proton-antiproton collisions at a center-of-mass energy of $\sqrt s =1.96$ TeV. We report the results of combinations of the inclusive asymmetries and their differential dependencies on relevant kinematic quantities. The combined inclusive asymmetry is $\afbtt= 0.128 \pm 0.025$. The combined inclusive and differential asymmetries are consistent with recent standard model predictions. 
\end{abstract}


\maketitle

The production of top and antitop quark ($t\bar{t}$) pairs at the Tevatron proton-antiproton ($p\bar{p}$) collider at Fermilab is dominated by the $q\bar{q}$ annihilation process, which can lead to asymmetries, $\afbtt$, in the number of top quarks produced within the hemisphere centered on the beam proton (forward) relative to those that are produced within the antiproton hemisphere (backward). In the standard model (SM), no forward-backward asymmetries are expected at leading order in perturbative quantum chromodynamics (QCD). However, contributions to the asymmetry from interference of leading order and higher-order amplitudes, and smaller offsetting contributions from the interference of initial- and final-state radiation, combine to yield a non-zero asymmetry~\cite{JHEP05(2016)034,Czakon:2014xsa,Bernreuther:2012sx,Kidonakis:2015,Brodsky:2017}. Compared to older predictions~\cite{oldSMRefs} of the inclusive asymmetry at next-to-leading order (NLO) QCD, the latest higher-order corrections in QCD and electroweak theory (EW) are almost of the same size as the inclusive prediction at NLO QCD. Measurements of the inclusive asymmetries and their dependence on kinematic quantities of top quarks and their decay leptons are used to probe the production mechanism. Beyond-the-SM (BSM) interactions \cite{BSMphysics} can significantly alter the dynamics, even such that differential asymmetries can be strikingly changed while inclusive asymmetries are only marginally affected. 

Inclusive and differential measurements \cite{oldAfbCDF, oldAfbD0} by the CDF~\cite{cdf-expt} and D0~\cite{d0-expt} Collaborations in 2011 were only marginally consistent with each other, and with then-existing SM predictions~\cite{oldSMRefs}. Both collaborations have since completed measurements using the full Tevatron Run II $\ppbar$ collision data, corresponding to integrated luminosities between 9 and 10 fb$^{-1}$. Assuming SM $t$ and $\bar{t}$ decays, they have measured asymmetries using events containing a single charged lepton ($\ell +$jets), where one $W$ boson from a top quark decays to a charged lepton and a neutrino and the other decays to a quark and an antiquark that evolve into jets, and in events containing two charged leptons ($\ell\ell$) where both $W$ bosons decay leptonically. Both collaborations have measured inclusive and differential asymmetries as functions of kinematic quantities of the top quarks and their decay leptons. More refined analysis techniques have been employed since the initial measurements. In the $\ell +$jets channel, CDF performed a detailed investigation of the inclusive and differential $\ttbar$ asymmetries \cite{Aaltonen:2012it}, and D0 used a novel partial event reconstruction for the inclusive and differential measurement of $\afbtt$ \cite{Abazov:2014cca}. In the $\ell\ell$ channel, CDF used several kinematic distributions to minimize the expected total uncertainty \cite{Aaltonen:2016bqv}, while D0 carried out a simultaneous measurement of $\afbtt$ and the top quark polarization \cite{D06445}. \\ 

We present the combinations of the final CDF and D0 measurements and compare them with current SM calculations~\cite{afb-rmp}. Careful assessment of the correlations of systematic uncertainties between analysis channels and experiments is required for comparing the data with predictions.

For reconstructed top and antitop quarks, $\afbtt$ is defined by
\begin{eqnarray}
\label{eqn:afbtt}
\afbtt =  \frac{N(\Dy > 0) - N(\Dy < 0)}{N(\Dy > 0) + N(\Dy < 0)} ~~,
\end{eqnarray}
where $\Dy = y_t - y_{\bar t}$ is the rapidity difference \cite{rapidity} between the $t$ and $\bar t$ quark, and $N$ is the signal yield in a particular configuration. Typically, measurements of $\ttbar$ forward-backward asymmetries require reconstruction of top and antitop quarks using all available information associated with the final-state particles \cite{kinReco}. Background contributions are subtracted from the yield of $\ttbar$ candidates, thereby providing the $\ttbar$ signal. The latter is corrected for detector effects, so as to unfold from the reconstructed $t$ and $\bar{t}$ quarks to the parton level. 

The asymmetry in $t$ and $\bar t$ quark production also leads to asymmetries in their decay leptons which, while smaller in magnitude, do not need unfolding, but must be corrected for acceptance effects. The single-lepton asymmetry is defined by
\begin{eqnarray}
\afblep = \frac{N(q_\ell \eta_\ell > 0) - N(q_\ell \eta_\ell < 0)}
{N(q_\ell \eta_\ell > 0) + N(q_\ell \eta_\ell < 0)} ~~,
\end{eqnarray}
where $q_\ell$ is the sign of the electric charge and $\eta_\ell$ the pseudorapidity of the lepton in the laboratory frame. For the $\ell\ell$ channel, the dilepton asymmetry is defined as
\begin{eqnarray}
\afbdeta = \frac{N(\Delta \eta > 0) - N(\Delta \eta < 0)}
{N(\Delta \eta > 0) + N(\Delta \eta < 0)} ~~,
\end{eqnarray}
where $\Delta \eta =\eta_{\ell^+} - \eta_{\ell^-}$ is the pseudorapidity difference between the positive- and negative-charge lepton.
The asymmetries obtained using top quarks and leptons are correlated, as a positive rapidity difference between a $t$ and a $\bar{t}$ quark is likely to produce a positive pseudorapidity difference between a positive- and negative-charge decay lepton.

Inclusive and differential measurements of $\afbtt$ at the Tevatron were reported in Refs.~\cite{Aaltonen:2012it, Abazov:2014cca} for the $\ell+$jets channel and in Refs.~\cite{Aaltonen:2016bqv, D06445} for the $\ell\ell$ channel. Measurements of $\afblep$ for the $\ell +$jets channel are given in Refs.~\cite{Aaltonen:2013vaf,Abazov:2014oea} and in Refs.~\cite{Aaltonen:2014eva,Abazov:2013wxa} for the $\ell\ell$ channel. Measurements of $\afbdeta$ are reported in Refs.~\cite{Aaltonen:2014eva,Abazov:2013wxa}.

We combine the following CDF and D0 results using the best linear unbiased estimator (BLUE)~\cite{Lyons1988,Lyons1990,Valassi}:
the inclusive asymmetries $\afbtt$, $\afblep$, and $\afbdeta$, each extrapolated to the full phase space relying on corresponding Monte Carlo simulations, and the differential asymmetry of $\afbtt$ as a function of the invariant mass of the $\ttbar$ system ($\Mtt$). For combinations of inclusive asymmetries the input uncertainties are symmetrized, while they are treated as asymmetric in the case of the combination of the asymmetry as a function of $\Mtt$. A mutually compatible classification of all systematic uncertainties is not available for $\afbtt$ as a function of $|\Dy|$. Hence, we provide results of a simultaneous least-squares fit to determine the slope parameter of the asymmetry in the CDF and D0 data, assuming a linear dependence. A similar fit is also provided for $\afbtt$ as a function of $\Mtt$. The CDF and D0 differential asymmetries, $\afblep$ as a function of $q_\ell \eta_\ell$ and $\afbdeta$ as a function of $\Delta \eta$ are not combined, but are displayed together for ease of comparison. 

Predictions of inclusive and differential $\afbtt$ distributions at next-to-next-to-leading order (NNLO) QCD calculations are available from Ref.~\cite{JHEP05(2016)034}. The contribution from EW NLO corrections to the NLO QCD asymmetries are not negligible~\cite{Bernreuther:2012sx}. Hence, we compare the measurements to the latest NNLO QCD + NLO EW  inclusive and differential $\afbtt$ calculations \cite{JHEP05(2016)034, CzakonEWcorrection:2017}. The combined inclusive-lepton asymmetries $\afblep$ and $\afbdeta$ are compared to the NLO QCD + NLO EW predictions of Ref.~\cite{Bernreuther:2012sx}.

To accommodate correlations among analysis channels and between experiments, we classify systematic uncertainties into the following categories: 

\begin{itemize}
	\item[(i)] {\it Background modeling}:  The uncertainties in the distribution and normalization of the background are assumed to be uncorrelated since the backgrounds are estimated differently in different analyses, and in the two experiments.
	\item[(ii)] {\it Signal modeling}:  The uncertainties in modeling the signal, parton showering \cite{PartonShower}, initial- and final-state radiation \cite{ISRFSR}, and color connections \cite{colorRe} are taken to be fully correlated among analysis channels and experiments because they all rely on the same assumptions. 
	\item[(iii)] {\it Detector modeling}:  The uncertainties in jet-energy scale \cite{JES} and the modeling of the detector are fully correlated within each experiment and uncorrelated between the two experiments. 
	\item[(iv)] {\it Method}:  The uncertainties in the methods used to correct for detector acceptance, efficiency, and potential biases in the reconstruction of top quark kinematic properties are mostly taken to be uncorrelated between experiments and analysis channels. However, the uncertainties on the phase-space correction procedures for the leptonic asymmetry in the D0 $\ell +$jets~\cite{Abazov:2014cca} and $\ell\ell$~\cite{D06445} analyses are estimated using the same methods and are therefore correlated with each other but are uncorrelated with the CDF results.
	\item[(v)] {\it PDF}:  The uncertainties in parton-density distribution functions (PDF) and the pileup in energy from overlapping $p\bar{p}$ interactions are treated as fully correlated between the analysis channels and the two experiments, because they characterize the same potential systematic biases.
\end{itemize}

The combined inclusive asymmetry is $\afbtt=
0.128 \pm 0.021 {\rm (stat)} \pm 0.014{\rm (syst)}$, consistent with the NNLO QCD + NLO EW prediction of $0.095\pm0.007$~\cite{Czakon:2014xsa} within 1.3 standard deviations (SD). The combination has a $\chi^2$ of $1.7$ for 3 degrees of freedom (dof). BLUE also provides the weights in the combination for the CDF $\ell+$jets, D0 $\ell+$jets, CDF $\ell\ell$, and D0 $\ell\ell$ results, which are 0.25, 0.64, 0.01, and 0.11, respectively. 

The CDF and D0 differential $\afbtt$ asymmetries as a function of $\Mtt$ are measured only for the $\ell +$jets channel. We combine the D0 bins in the range of $350 < \Mtt < 550$ $\gevcc$ to provide uniform, 100 $\gevcc$-wide, bins for the combination. For the two measurements we use covariance matrices \cite{epaps} that take into account the bin-to-bin correlations from the unfolding of differential distributions. The correlations in systematic uncertainties among channels and experiments for each $\Mtt$ bin are assumed to be equal to those in the inclusive measurements. However, the uncorrelated background uncertainties for the differential asymmetries are subdivided into two separate components, one for the overall normalization and one for the differential distribution (shape) of the background. According to the different experimental methodologies, these are treated as correlated between bins for the CDF measurement and as uncorrelated for the D0 measurement. We verify that changing the correlations of systematic uncertainties between $-1$ and $+1$ has negligible impact on the combined result because the statistical uncertainties dominate.

\begin{figure}[hbt]
\includegraphics[width=0.485\textwidth]{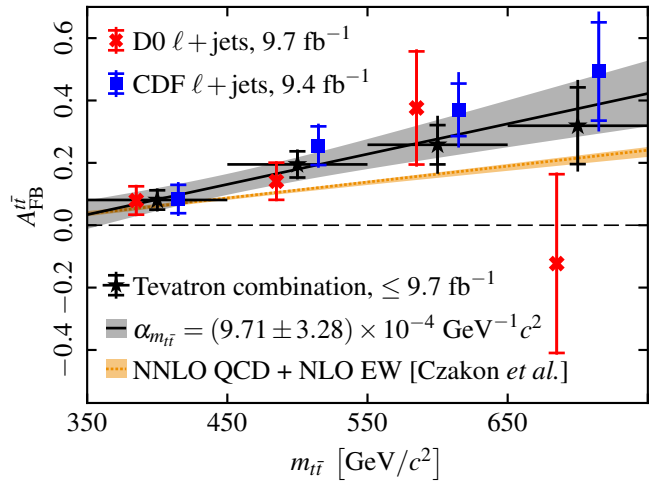}
\caption{Results for $\afbtt$ vs.\ $\Mtt$ for the individual CDF and D0 measurements and for their combination. The inputs to the combination are displaced at different abscissa values within each $\Mtt$ bin for ease of visibility. The inner error bar indicates the statistical uncertainty, while the outer error bar corresponds to the total uncertainty including the systematic uncertainty added in quadrature. The value of the combined data point for the mass region of $550 - 650~\gevcc$ is discussed in Ref.\ \cite{epaps} in more detail. The linear dependence of the combined result is given by the solid black line together with the 1 SD total uncertainty of the two-parameter fit given by the shaded gray area. The dashed orange line shows the NNLO QCD + NLO EW prediction of Refs.~\cite{JHEP05(2016)034,Czakon:2014xsa, CzakonEWcorrection:2017}, while the shaded orange area reflects its 1 SD uncertainty. 
}
\label{fig:AFBtt-mtt-combo}
\end{figure}

The combined $\afbtt$ values, and their statistical and systematic uncertainties for each category,  are given in Table~\ref{tab:AFBtt-mass-uncert}, which also reports the probabilities for the CDF and D0 inputs to agree with each other in each mass bin. Overall, the differential combination has a $\chi^2$ of $5.2$ for 4 dof. The correlations in the total uncertainties between $\Mtt$ bins are given in Ref.~\cite{epaps}. The values of $\afbtt$ as a function of $\Mtt$ for each experiment and their combination are shown in Fig.~\ref{fig:AFBtt-mtt-combo}, together with the NNLO QCD + NLO EW predictions \cite{CzakonEWcorrection:2017}. 

The counter-intuitive value of the combined asymmetry in the $550 - 650~\gevcc$ mass bin is due to the specific pattern of the CDF and D0 bin-to-bin correlations stemming from different choices in the regularized matrix unfolding. The opposite correlations observed between the $600~\gevcc$ and the $700~\gevcc$ mass bins in the CDF (large and positive) and D0 (small and negative) measurements give rise to a combined asymmetry in the $600~\gevcc$ bin that is smaller than that found in either measurement \cite{epaps}. 

To reduce the correlations between the slope and the intercept, we use a linear fit of the form $\afbtt(\Mtt) = \alpha_{\Mtt} (\Mtt - 450~\gevcc) + \beta_{\Mtt}$. The linear fit yields a slope of $\alpha_{\Mtt} = (9.71 \pm 3.28) \times 10^{-4}~\mathrm{GeV}^{-1}c^2$ with an intercept at a $\Mtt$ value of $450~\gevcc$ of $\beta_{\Mtt} = 0.131 \pm 0.034$. The fit has a $\chi^2$ of $0.3$ for 2 dof. The values predicted at NNLO QCD + NLO EW are $\alpha_{\Mtt}^{\rm SM} =\left(5.11 ^{+0.42}_{-0.64}\right) \times 10^{-4}~\mathrm{GeV}^{-1}c^2$ and an intercept of $\beta_{\Mtt}^{\rm SM} = 0.087 ^{+0.005}_{-0.006}$. The predicted dependence is determined by a linear fit to the binned prediction from Ref.\ \cite{CzakonEWcorrection:2017}. The NNLO QCD + NLO EW predictions of the differential $\afbtt$ and of the slope parameters agree with the combined experimental results to within 1.3 SD. 

\begin{table*}[htbp]
\begin{center}
\caption{Combined differential $\afbtt$ values in bins of $\Mtt$, with the probability (Prob) for the CDF and D0 inputs to agree with each other, with statistical (Stat), systematic (Tot syst), and total uncertainties. The systematic uncertainties are broken down into uncertainties in the distribution of the background (Bkd distr), background normalization (Bkd norm), signal modeling (Signal), detector modeling (Det), measurement method (Meth), and parton distribution function (PDF). 
}
\label{tab:AFBtt-mass-uncert}
\begin{ruledtabular}
\begin{tabular}{cccccccccccc}
\multirow{2}{*}{$\Mtt$ ($\gevcc$)}& \multirow{2}{*}{~~~$\afbtt$~~~} &\multirow{2}{*}{Prob}&\multicolumn{9}{c}
{Uncertainty}\\ \cline{4-12}
                         &&& Total  & Stat & Meth & Signal & PDF & Det & Bkd distr & Bkd norm & Tot syst \\\hline
350--450 & 0.081 & 95\%  & 0.037 &0.031 & 0.009  &0.012 & 0.004 & 0.007 &0.010 &0.003 & 0.020     \\
450--550 & 0.195 & 22\%  & 0.048 & 0.042 & 0.010 &0.016 & 0.007 & 0.006  &0.007 &0.006 & 0.023   \\
550--650 & 0.258 & 98\%  & 0.093 & 0.063 &0.008  &0.062 & 0.017 & 0.017  &0.006 & 0.008 & 0.068   \\
$> 650$ & 0.319 &    8\%  & 0.147 & 0.123 & 0.018 &0.065 & 0.021 & 0.026  &0.019 &0.019  & 0.080  \\
\end{tabular}
\end{ruledtabular}
\end{center}
\end{table*}

\begin{figure}[hbt]
\includegraphics[width=0.485\textwidth]{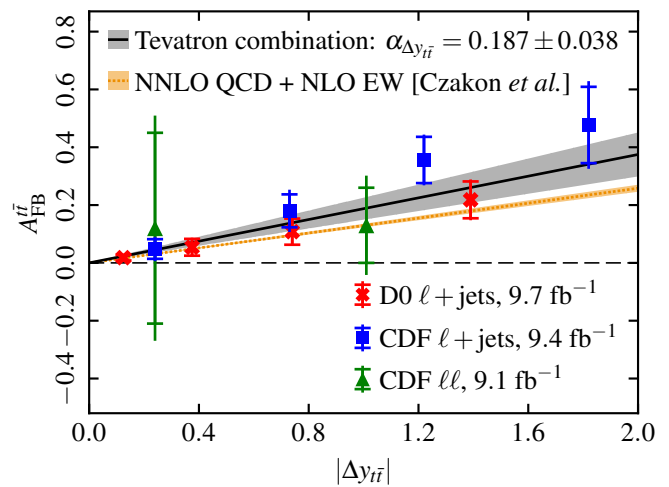}
\caption{Measurements of the differential asymmetries $\afbtt$ vs.\ $|\Dy|$ with data points displayed at the distribution-weighted center of the bins. The inner error bar indicates the statistical uncertainty, while the outer error bar corresponds to the total uncertainty, including the systematic uncertainty added in quadrature. The combined linear dependence for all the experimental results is given by the solid black line, with the 1 SD total uncertainty on the one-parameter fit given by the shaded gray area. The dashed orange line shows the NNLO QCD + NLO EW prediction~\cite{JHEP05(2016)034,Czakon:2014xsa, CzakonEWcorrection:2017}, while the shaded orange area reflects its 1 SD uncertainty.}
\label{fig:AFBtt-dy-combo}
\end{figure}

The differential $t\bar t$ asymmetry as a function of $|\Dy|$ is available from CDF for both the $\ell +$jets and $\ell\ell$ channels, and from D0 for the $\ell +$jets channel. The choice of binning differs for these measurements. We perform a simultaneous least-squares fit to a linear function $\afbtt(|\Dy|) = \alpha_{\Dy} |\Dy|$ for all available measurements, employing a combined $10 \times 10$ covariance matrix ${\cal C}_{ij}$. We define $\chi^2(|\Dy|) = \sum_{ij} [y_i - f_i(|\Dy|)]~{\cal C}_{ij}^{-1}~[y_j - f_j(|\Dy|)]$, with $y_{i}$ and $y_{j}$ representing the bin $i$ and $j$ of each of the three measurements, and $f_{i}(|\Dy|)$ and $f_{j}(|\Dy|)$ representing the expectations from a linear function. The definition of the asymmetry ensures that $\afbtt=0$ at $\Dy =0$. The correlations of the systematic uncertainties among analysis channels and experiments are assumed to be equal to those in the $\afbtt$ vs.\ $\Mtt$ measurements. Figure ~\ref{fig:AFBtt-dy-combo} shows the individual measurements and the result of the linear fit. The linear dependence for the combination is measured to be $\alpha_{\Dy} = 0.187 \pm 0.038$ with a $\chi^2$ of 10.9 for 9 dof. A fit to the binned NNLO QCD + NLO EW predictions of Ref.~\cite{JHEP05(2016)034,Czakon:2014xsa, CzakonEWcorrection:2017} gives the slope $\alpha_{\Dy}^{\rm SM} = 0.129^{+0.006}_{-0.012}$. The prediction and the combined result differ by 1.5 SD. 

\begin{figure}[hbt]
\includegraphics[width=0.485\textwidth]{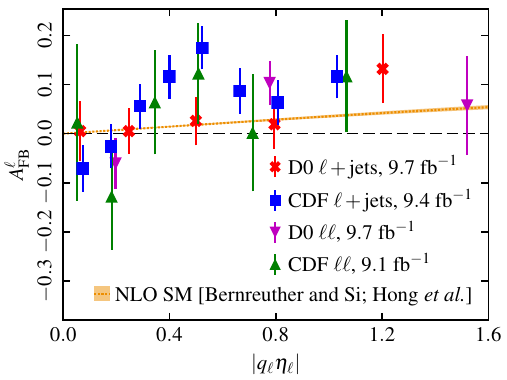}
\caption{Comparison of the differential asymmetries $\afblep$ as a function of  $|q_{\ell} \eta_{\ell}|$. Each error bar represents the total experimental uncertainty. The dashed orange line shows the NLO SM prediction~\cite{Bernreuther:2012sx, Hong:2014}, while the shaded orange area shows its 1 SD uncertainty.
}
\label{fig:AFBlepDiff}
\end{figure}

The combined fit to the CDF and D0 inclusive single-lepton asymmetries gives $\afblep=0.073 \pm 0.016 {\rm (stat)} \pm 0.012{\rm (syst)}$.  The fit has a $\chi^2$ of $2.2$ for 3 dof, and the result is consistent with the NLO QCD + NLO EW prediction of $0.038\pm0.003$~\cite{Bernreuther:2012sx} to within 1.6 SD. The weights of the CDF $\ell+$jets, D0 $\ell+$jets, CDF $\ell\ell$ and D0 $\ell\ell$ results in the fit are 0.40, 0.27, 0.11, and 0.23, respectively. The individual CDF and D0 measurements of $\afblep$ as a function of $q_\ell \eta_\ell$ are shown in Fig.~\ref{fig:AFBlepDiff}.

\begin{figure}[hbt]
\includegraphics[width=0.485\textwidth]{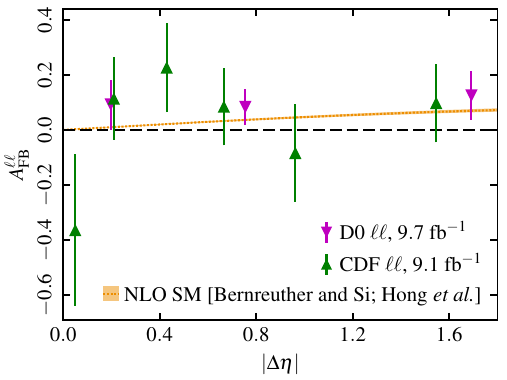}
\caption{Comparison of the differential asymmetries $\afbdeta$ as a function of $|\deta|$. Each error bar represents the total experimental uncertainty. The dashed orange line shows the NLO SM  prediction~\cite{Bernreuther:2012sx, Hong:2014}, while the shaded orange area shows its 1 SD uncertainty.
}
\label{fig:AFBdetaDiff}
\end{figure}

The combined fit to the CDF and D0 inclusive $\afbdeta$ measurements yields $\afbdeta=0.108\pm 0.043 {\rm (stat)} \pm 0.016{\rm (syst)}$.  The fit has a $\chi^2$ of $0.2$ for 1 dof, and the result is consistent with the NLO QCD + NLO EW prediction of $0.048\pm0.004$~\cite{Bernreuther:2012sx} to within 1.3 SD. The weights of the CDF and D0 $\ell\ell$ results in the fit are 0.32 and 0.68, respectively. The individual CDF and D0 measurements of $\afbdeta$ as a function of $\Delta\eta$ are shown in Fig.~\ref{fig:AFBdetaDiff}.

\begin{figure}[hbt]
\includegraphics[width=0.485\textwidth]{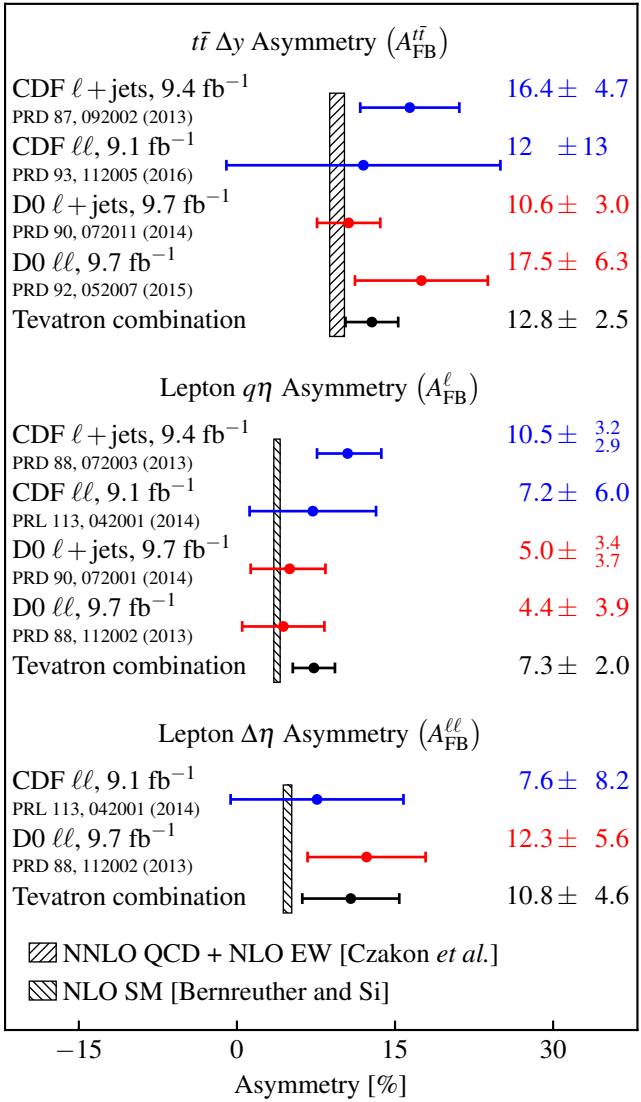}
\caption{Summary of inclusive forward-backward asymmetries in $\ttbar$ events in percents at the Tevatron. 
}
\label{fig:AFBTev}
\end{figure}

In summary, we report combinations of the measurements of top-antitop quark forward-backward asymmetries performed in a $\ppbar$ collision sample corresponding to 9$-$10~fb$^{-1}$ collected by the CDF and D0 experiments at the Tevatron. Both the inclusive and differential measurements favor somewhat larger positive asymmetries than the predictions. The resulting combined inclusive asymmetry is $\afbtt= 0.128 \pm 0.025$ compared to the prediction at NNLO QCD + NLO EW of $0.095\pm0.007$. All three inclusive observables agree with the existing SM predictions to within 1.6 standard deviations. The differential asymmetries as a function of $\Mtt$ and $\Dy$ agree to within 1.5 standard deviations. We conclude that the measurements and their combinations, shown in Fig.\ \ref{fig:AFBTev}, are consistent with each other and with the SM predictions. The reported consistency is the result of an intense effort of refining the experimental and theoretical understanding, which started in 2010, when significant departures of the first Tevatron measurements \cite{oldAfbCDF, oldAfbD0} from the predictions suggested potential contributions from BSM dynamics.
 
This document was prepared by the CDF and D0 collaborations using the resources of the Fermi National Accelerator Laboratory (Fermilab), a U.S. Department of Energy, Office of Science, HEP User Facility. Fermilab is managed by Fermi Research Alliance, LLC (FRA), acting under Contract No. DE-AC02-07CH11359.  

We thank the staffs at Fermilab and collaborating institutions, and acknowledge support from
the Department of Energy and
the National Science Foundation (U.S.A.), 
the Australian Research Council (Australia), 
the National Council for the
Development of Science and Technology and 
the Carlos Chagas Filho Foundation for the Support of Research in the
State of Rio de Janeiro (Brazil), 
the Natural Sciences and Engineering Research Council (Canada), 
the China Academy of Sciences,
 the National Natural Science
Foundation of China, and 
the National Science Council
of the Republic of China (China), 
the Administrative Department of Science, Technology and Innovation (Colombia), 
the Ministry of Education, Youth and Sports (Czech Republic), 
the Academy of Finland, 
the Alternative Energies and Atomic Energy Commission and 
the National Center for Scientific Research/National Institute of Nuclear and Particle Physics (France), 
the Bundesministerium f\"ur Bildung und Forschung (Federal Ministry of Education and Research) and 
the Deutsche Forschungsgemeinschaft (German Research Foundation) (Germany), 
the Department of Atomic Energy and Department of Science and Technology (India), 
the Science Foundation Ireland (Ireland), 
the National Institute for Nuclear Physics (Italy), 
the Ministry of Education, Culture, Sports, Science and Technology (Japan), 
the Korean World Class University Program and 
the National Research Foundation of Korea (Korea), 
the National Council of Science and Technology (Mexico), 
the Foundation for Fundamental Research on Matter (Netherlands), 
the Ministry of Education and Science of the Russian Federation, 
the National Research Center "Kurchatov Institute" of the Russian Federation, and 
the Russian Foundation for Basic Research (Russia), 
the Slovak R\&D Agency (Slovakia), 
the Ministry of Science and Innovation, and 
the Consolider-Ingenio 2010 Program (Spain), 
the Swedish Research Council (Sweden), 
the Swiss National Science Foundation (Switzerland), 
the Ministry of Education and Science of Ukraine (Ukraine),
the Science and Technology Facilities Council and  
The Royal Society (United Kingdom), 
the A. P. Sloan Foundation (U.S.A.), and 
the European Union community Marie Curie Fellowship Contract No. 302103.

\clearpage

\appendix
\onecolumngrid
\section{\label{sec:app} Appendix: Supplemental material}
In this appendix, we provide supplemental information on the combination of the CDF and D0 measurements of the forward-backward asymmetries in $t\overline t$ pair production at the Fermilab Tevatron. \\ 

\section{\label{sec:afbtt} $t\overline t$ production asymmetry, $\afbtt$}

Table~\ref{tab:AFBttUncertainties} reports the uncertainties for each of the contributing measurements to the inclusive $t\overline t$ asymmetry, $\afbtt$, and the uncertainties for their combination in the fit. Table~\ref{tab:AFBttResults} shows the individual inclusive $\afbtt$ measurements and uncertainties, as well as their combination. The contribution, in terms of the weights determined by BLUE \cite{Lyons1988,Lyons1990,Valassi}, of each measurement in the fit are also shown.\\

\begin{figure}[hbt]
\includegraphics[width=0.585\columnwidth]{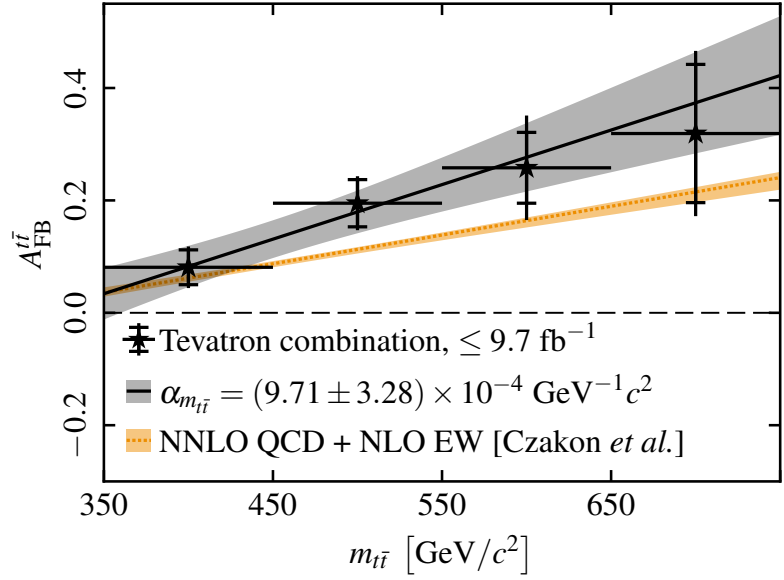}
\caption{Differential $\afbtt$ vs.\ $\Mtt$ for the Tevatron combination. The linear slope of the combined result is given by the solid black line together with the total uncertainty of the two-parameter fit (shaded gray area). The dashed solid orange line shows the NNLO QCD + NLO EW prediction of Refs.~\cite{JHEP05(2016)034,Czakon:2014xsa, CzakonEWcorrection:2017}, while the shaded orange area shows the 1 SD theoretical uncertainty on the prediction.}
\label{fig:AFBttMttDiff_slope}
\end{figure}

\begin{figure}[hbt]
\includegraphics[width=0.585\columnwidth]{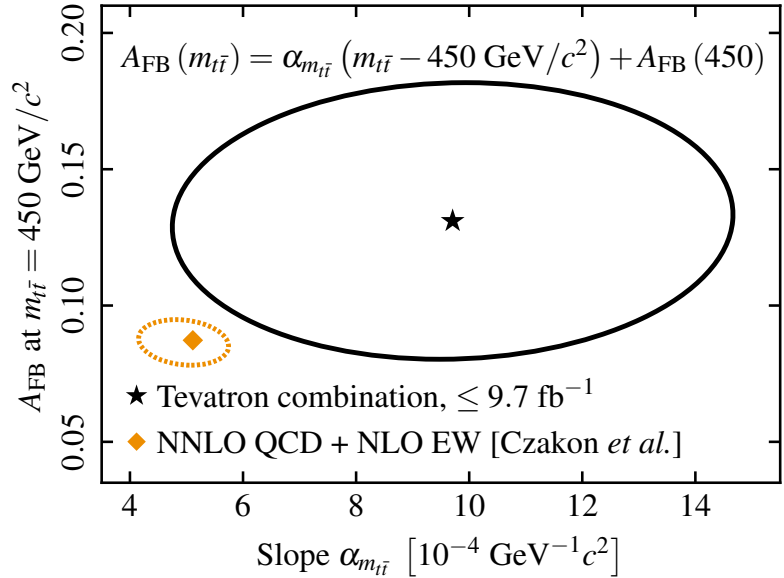}
\caption{Correlation of the slope and intercept from the linear fit of the $\afbtt$ vs.\ $\Mtt$ data represented as 68\% confidence ellipses, and shown at $\Mtt = 450$ $\gevcc$. The smaller dashed orange ellipse shows the correlation of the slope and intercept of the NNLO QCD + NLO EW prediction of Refs.~\cite{JHEP05(2016)034,Czakon:2014xsa, CzakonEWcorrection:2017}.}
\label{fig:AFBttMttDiff_slopeCorr}
\end{figure}

Table~\ref{tab:afb_mtt_systs} shows the inputs to the differential $\afbtt$ vs.\ $\Mtt$ fit and their uncertainties. Figure~\ref{fig:AFBttMttDiff_slope} shows the combined result for the differential $\afbtt$ vs.\ $\Mtt$ data. The linear fit to the data and its one standard deviation (SD) uncertainty are shown by the black solid line and gray shaded band; the corresponding quantities for the theoretical prediction are shown by the orange line and shaded band. Figure~\ref{fig:AFBttMttDiff_slopeCorr} shows the correlations of the slope and intercept at $\Mtt = 450$ $\gevcc$ for the data and theoretical prediction. The smaller orange ellipse shows the correlation of the slope and intercept of the NNLO QCD + NLO EW prediction of Ref.~\cite{JHEP05(2016)034, Czakon:2014xsa, CzakonEWcorrection:2017}. 

\begin{figure}[hbtp]
\includegraphics[width=0.495\columnwidth,angle=0]{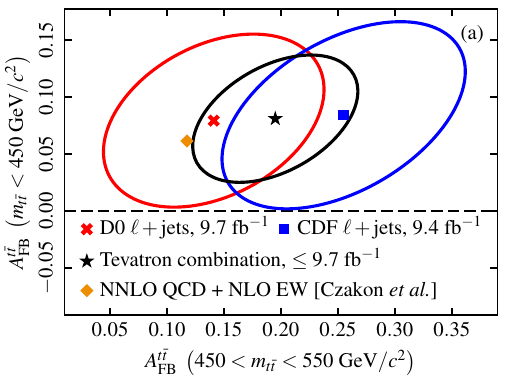}
\includegraphics[width=0.495\columnwidth,angle=0]{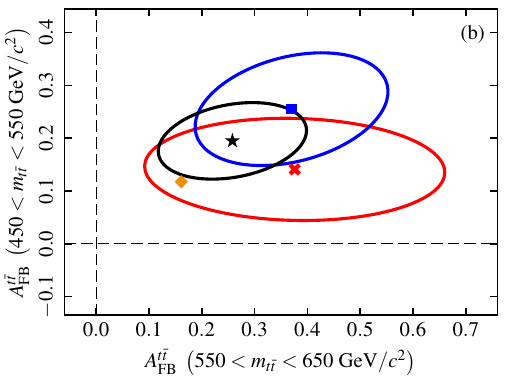}
\includegraphics[width=0.495\columnwidth,angle=0]{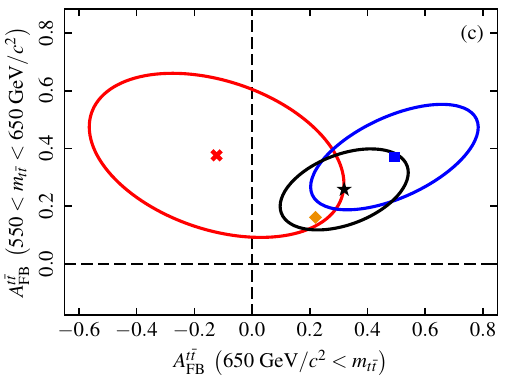}
\caption{Correlations of combined statistical and systematic uncertainties represented as 68\% confidence ellipses between the first and second $\afbtt$ vs.\ $\Mtt$ bin (a), the second and third $\afbtt$ vs.\ $\Mtt$ bin (b), and the third and fourth $\afbtt$ vs.\ $\Mtt$ bin (c). The orange marker shows the NNLO QCD + NLO EW theoretical prediction~\cite{JHEP05(2016)034,Czakon:2014xsa, CzakonEWcorrection:2017} while the dashed orange ellipse shows the theoretical uncertainty on the prediction.}
\label{fig:AFBttMttDiff_corrs}
\end{figure}

Figure~\ref{fig:AFBttMttDiff_corrs} shows the correlations of total uncertainties between adjacent $\Mtt$ bins, for the CDF and D0 data, as well as for the combination. The correlations between the individual CDF and D0 measurements for the third bin with adjacent bins of $\Mtt$ result in a combined asymmetry value that is smaller than either of the inputs for the third bin. This behavior indicates the presence of large correlations and can be understood when looking at the orientation of the correlation ellipses in Figure~\ref{fig:AFBttMttDiff_corrs}(c). The 68\% confidence level ellipses show smaller uncertainties for the CDF inputs than for the D0 inputs. The smaller CDF uncertainties results from the different choice made by CDF for the regularization method used to correct for detector effects. 

Table~\ref{tab:afbTeV_mtt_covs} shows the covariance matrix of the combined differential $\afbtt$ vs.\ $\Mtt$. Table ~\ref{tab:afb_dy_systs} shows the inputs to the differential $\afbtt$ vs.\ $|\Dy|$ fit and their uncertainties. Table~\ref{tab:covarianceMatdY} shows the covariance matrix of total uncertainties of the differential $\afbtt$ vs.\ $|\Dy|$ inputs to the combined fit. 

\section{\label{sec:afblep}Single lepton asymmetry, $\afblep$}

Table~\ref{tab:AFBlepUncertainties} reports the uncertainties for each of the contributing measurements to the inclusive single lepton asymmetry, $\afblep$, and the uncertainties for their combination in the fit. Table~\ref{tab:AFBlepResults} shows the individual inclusive $\afblep$ measurements and uncertainties, as well as their combination. The weights of each measurement in the fit are also shown. 

\section{\label{sec:afbdilep}Dilepton asymmetry, $\afbdeta$}

Table~\ref{tab:AFBdetaUncertainties} reports the uncertainties for each of the contributing measurements to the inclusive dilepton asymmetry $\afbdeta$ and the uncertainties for their combination in the fit. Table~\ref{tab:AFBdetaResults} shows the individual inclusive $\afbdeta$ measurements and uncertainties, as well as their combination. The weights of each measurement in the fit are also shown.

\vspace*{2.0cm}

\begin{table*}[htbp]
\begin{center}
\caption{Statistical and systematic uncertainties in the individual inclusive $\afbtt$ inputs as well as in the resultant combination.}
\begin{ruledtabular}
\label{tab:AFBttUncertainties}
\begin{tabular}{lccccc}
Uncertainty & CDF $\ell$+jets \cite{Aaltonen:2012it} & CDF $\ell\ell$ \cite{Aaltonen:2016bqv} & D0 $\ell$+jets \cite{Abazov:2014cca} & D0 $\ell\ell$ \cite{D06445} & Combination \\\hline
Statistical & 0.039 & 0.11 & 0.027 & 0.056 & 0.021\\
Background & 0.022 & 0.04 & 0.010 & 0.007&  0.008\\
Signal & 0.011 & 0.05  & 0.005 & 0.026 & 0.009\\
Detector & 0.007  & 0.02 & 0.003 & 0.001 & 0.003\\
Method & 0.004 & 0.02  & 0.005 & 0.014 & 0.004\\
PDF & 0.001 & 0.01 & 0.004 & 0.003 & 0.003\\
\end{tabular}
\end{ruledtabular}
\end{center}
\end{table*}

\begin{table*}[htbp]
\begin{center}
\caption{Inputs to the combination of the inclusive $t\bar t$ asymmetries and results of the combination. 
}
\begin{ruledtabular}
\label{tab:AFBttResults}
\begin{tabular}{lccccc}
\multirow{2}{*}{Analysis}& \multirow{2}{*}{$\afbtt$} &\multicolumn{3}{c}{Uncertainty}&\multirow{2}{*}{Weight}\\\cline{3-5}
&& Stat. & Syst. & Total&\\\hline
CDF $\ell+$jets \cite{Aaltonen:2012it} & 0.164& 0.039& 0.026& 0.047& 0.25\\
CDF $\ell\ell$ \cite{Aaltonen:2016bqv} & 0.12 & 0.11 & 0.07 & 0.13 & 0.01\\
D0 $\ell+$jets \cite{Abazov:2014cca} & 0.106 & 0.027 & 0.013 & 0.030 & 0.64\\
D0 $\ell\ell$ \cite{D06445} & 0.175 & 0.056 & 0.031 & 0.063 & 0.11 \\\hline
Combination& 0.128 & 0.021 & 0.014 & 0.025 & \\
\end{tabular}
\end{ruledtabular}
\end{center}
\end{table*}

\begin{table*}[hbtp]
\caption{Inputs of the $\afbtt$ results in the $\ell +$jets channels, along with their statistical (Stat) and systematic uncertainties broken down for the individual $\Mtt$ bins. The listed systematic uncertainties originate from the measurement method (Method), signal modeling (Signal), parton-distribution function (PDF), detector modeling (Detector), and from background shape (Bkd dist) and background normalization (Bkd norm).}
\label{tab:afb_mtt_systs}
\begin{ruledtabular}
\begin{tabular}{lccccccccc}
 \multirow{2}{*}{$\Mtt$ [$\gevcc$]} & \multirow{2}{*}{$\afbtt$} &\multicolumn{8}{c}{Uncertainty}\\ \cline{3-10} 
 & &Total & Stat & Method & Signal & PDF & Detector & Bkd dist & Bkd norm \T \\ \hline 
\underline{D0 $\ell +$jets} & & & & & & & & &  \\
 350--450 & $\hphantom{+}$0.079 & 0.050 & 0.046 & 0.011 &  0.015 &  0.005 &  0.005 &  0.007 &  0.001\\
 450--550 & $\hphantom{+}$0.141 & 0.064 & 0.060 & 0.018 &  0.010 &  0.011 & 0.0024 &  0.007 &  0.001\\
 550--650 & $\hphantom{+}$0.376 &  0.188 &0.181 & 0.011 & 0.028 &  0.035 &  0.018 &  0.010 &  0.002\\
 $>650$ & $-0.123$ &  0.292 &0.287 & 0.017 &  0.009 &  0.043 &  0.030 &  0.014 &  0.003\\ \hline

\underline{CDF $\ell +$jets} & & & & & & & & &  \\
 350--450 & $\hphantom{+}$0.084 &  0.055 &  0.046 &  0.012 &  0.009 &  0.001 &  0.013 &  0.021 &  0.008\\
 450--550 & $\hphantom{+}$0.255 & 0.071 & 0.062 & 0.002 & 0.021 & 0.001 & 0.013 & 0.017 &  0.016\\
 550--650 & $\hphantom{+}$0.370 &  0.121 &  0.084 & 0.001 &  0.077 & 0.001 & 0.032 &  0.011 &  0.021\\
 $>650$ & $\hphantom{+}$0.493 & 0.193 & 0.158 & 0.023 & 0.091 & 0.001 & 0.045 & 0.021 & 0.031\\ 
\end{tabular}
\end{ruledtabular}
\end{table*}

\begin{table*}[hbtp]
\caption{Covariance matrix of the combined CDF and D0 differential $\afbtt$ vs.\ $\Mtt$.}
\label{tab:afbTeV_mtt_covs}
\begin{ruledtabular}
\begin{tabular}{lcccc}
$\Mtt$ [$\gevcc$] & 350--450 & 450--550 & 550--650 & $>650$ \\\cline{1-5}
350--450 & +0.0013690 &+0.0007672   &+0.0002512  &+0.0003644 \\
450--550 & +0.0007672 &+0.0023040   &+0.0012140  &$-$0.0005292 \\
550--650 & +0.0002512 &+0.0012140   &+0.0086490  &+0.0057140 \\
$>650$ &   +0.0003644 &$-$0.0005292&+0.0057140  &+0.0216100  \\ 
\end{tabular}
\end{ruledtabular}
\end{table*}

\begin{table*}[hbtp]
\caption{Inputs of the differential $\afbtt$ vs.\ $|\Dy|$ results in the $\ell+$jets channels and $\ell\ell$ channel. Their statistical and systematic uncertainties are broken down for the individual $|\Dy|$ bins. The listed systematic uncertainties originate from the measurement method (Method),  signal modeling (Signal), parton-distribution function (PDF), detector modeling (Detector), and from background shape (Bkd dist) and background normalization (Bkd norm).}
\label{tab:afb_dy_systs}
\begin{ruledtabular}
\begin{tabular}{lccccccccc}
\multirow{2}{*}{$\Dy$}& \multirow{2}{*}{$\afbtt$} &\multicolumn{8}{c}{Uncertainty}\\ \cline{3-10} 
&&Total & Statistical & Method & Signal & PDF & Detector & Bkd dist & Bkd norm \\ \hline
\underline{D0 $\ell +$jets} & & & & & & & & &  \\
 0.00--0.25 & 0.018 & 0.012 & 0.010 & 0.004 & 0.004 &  0.002 &  0.003 &  0.004 &  0.001 \\
 0.25--0.50 & 0.054 & 0.033 & 0.029 & 0.009 & 0.008 &  0.003 &  0.005 &  0.008 & 0.001 \\
 0.50--1.00 & 0.108 &  0.048 &  0.045 & 0.010 & 0.009 &  0.004 &  0.006 &  0.009 &  0.001 \\
 $> 1.00$ & 0.218 &  0.071 &  0.064 & 0.017&  0.015 &  0.007 &  0.010 &  0.016 &  0.002 \\ \hline

\underline{CDF $\ell +$jets} & & & & & & & & &  \\
 0.00--0.50 & 0.048 &  0.042 &  0.034 &  0.004 &  0.017 &  0.001 & 0.005 &  0.017 &  0.005 \\
 0.50--1.00 & 0.180 &  0.074 &  0.057 &  0.008 & 0.027 &  0.001 &  0.015 &  0.029 &  0.017 \\
 1.00--1.50 & 0.356 &  0.088 &  0.080 &  0.001 &  0.013 &  0.001 &  0.004 &  0.005 &  0.032 \\
 $>1.50$ & 0.477 &  0.151 &  0.132 &  0.018 &  0.034 & 0.004 &  0.012 &  0.044 &  0.043 \\ \hline

\underline{CDF $\ell \ell$} & & & & & & & & &  \\
 0.00--0.50 & 0.12$\hphantom{0}$ &  0.39$\hphantom{0}$  &  0.33$\hphantom{0}$  &  0.06$\hphantom{0}$  &  0.16$\hphantom{0}$  &  0.01$\hphantom{0}$  &  0.02$\hphantom{0}$  &  \multicolumn{2}{c}{0.13} \\
 $>0.50$ & 0.13$\hphantom{0}$  &  0.17$\hphantom{0}$  &  0.13$\hphantom{0}$  &  0.02$\hphantom{0}$  &  0.09$\hphantom{0}$  &  0.01$\hphantom{0}$  &  0.02$\hphantom{0}$  &  \multicolumn{2}{c}{0.06} \\ 
\end{tabular}
\end{ruledtabular}
\end{table*}

\begin{sidewaystable}[hbtp]
\footnotesize
\caption{Covariance matrix of the statistical and systematic uncertainties of the differential $\afbtt$ vs.\ $|\Dy|$ results employed in the combined $\chi^2$ fit. The PDF and signal uncertainties are assumed to be fully correlated between CDF and D0 ($+1$), while others are assumed to be uncorrelated. }
\label{tab:covarianceMatdY}
\begin{ruledtabular}
\begin{tabular}{lcccccccccc}
& \multicolumn{4}{c}{D0 $\ell +$jets} & \multicolumn{4}{c}{CDF $\ell +$jets} & \multicolumn{2}{c}{CDF $\ell \ell$} \\ \cline{2-5} \cline{6-9} \cline{10-11}
 $\Dy$ & 0.00--0.25 & 0.25--0.50 & 0.50--1.00 & $>1.00$ & 0.00--0.50 & 0.50--1.00 & 1.00--1.50 & $>1.50$ & 0.00--0.50 & $>0.50$ \\ \cline{1-11}
\underline{D0 $\ell +$jets} & & & & & & & & & & \\ 
 0.00--0.25 & +1.5590e-04 & 	+3.1260e-04 & 	+4.3580e-04 & 	+2.0598e-04 & 	+6.7363e-05 & 	+1.0965e-04 & 	+5.4530e-05 & 	+1.4314e-04 & 	+6.1059e-04 & +3.8000e-04 \\
 0.25--0.50 & +3.1260e-04 & 	+1.0914e-03 & 	+1.3690e-03 & 	+5.2010e-04 & 	+1.3438e-04 & 	+2.1827e-04 & 	+1.0809e-04 & 	+2.8259e-04 & 	+1.2112e-03 & 	+7.5000e-04 \\ 
 0.50--1.00 & +4.3580e-04 & 	+1.3690e-03 & 	+2.3518e-03 & 	+1.1144e-03 & 	+1.5107e-04 & 	+2.4518e-04 & 	+1.2126e-04 & 	+3.1653e-04 & 	+1.3588e-03 & 	+8.4000e-04  \\ 
 $>1.00$ & +2.0598e-04 & 	+5.2010e-04 & 	+1.1144e-03 & 	+4.9802e-03 & 	+2.5234e-04 & 	+4.1067e-04 & 	+2.0392e-04 & 	+5.3492e-04 & 	+2.2847e-03 & 	+1.4200e-03  \\ \hline
\underline{CDF $\ell +$jets} & & & & & & & & & & \\ 
 0.00--0.50 & +6.7363e-05 & 	+1.3438e-04 & +1.5107e-04 & 	+2.5234e-04 & 	+1.7632e-03 & 	+2.0343e-03 & 	+2.0723e-05 & 	$-$2.0149e-03 & 	+2.5367e-03 & 	$-$1.3013e-03 \\ 
 0.50--1.00 & +1.0965e-04 & 	+2.1827e-04 & 	+2.4518e-04 & 	+4.1067e-04 & 	+2.0343e-03 & 	+5.3518e-03 & 	+2.4451e-03 & 	$-$2.7678e-04 & 	$-$1.4144e-04 & 	+1.1556e-04  \\ 
 1.00--1.50 & +5.4530e-05 & 	+1.0809e-04 & 	+1.2126e-04 & 	+2.0392e-04 & 	+2.0723e-05 & 	+2.4451e-03 & 	+7.6970e-03 & 	+1.1443e-02 & 	$-$5.2223e-04 & 	+6.4377e-04  \\ 
 $>1.50$ &1.4314e-04 & 	+2.8259e-04 & 	+3.1653e-04 & 	+5.3492e-04 & 	$-$2.0149e-03 & 	$-$2.7678e-04 & 	+1.1443e-02 & 	+2.2995e-02 & 	$-$3.0824e-03 & 	+2.6086e-03  \\ \hline
\underline{CDF $\ell \ell$} & & & & & & & & & & \\
0.00--0.50 & +6.1059e-04 & 	+1.2112e-03 & 	+1.3588e-03 & 	+2.2847e-03 & 	+2.5367e-03 & 	$-$1.4144e-04 & 	$-$5.2223e-04 & 	$-$3.0824e-03 & 	+1.5170e-01 & 	$-$2.2280e-02 \\ 
$>0.50$ & +3.8000e-04 & 	+7.5000e-04 & 	+8.4000e-04 & 	+1.4200e-03 & 	$-$1.3013e-03 & 	+1.1556e-04 & 	+6.4377e-04 & 	+2.6086e-03 & 	$-$2.2280e-02 & 	+2.9500e-02 \\ 
\end{tabular}
\end{ruledtabular}
\end{sidewaystable}

\begin{table*}[htbp]
\begin{center}
\caption{Statistical and systematic uncertainties in the individual inclusive $\afblep$ inputs. 
}
\label{tab:AFBlepUncertainties}
\begin{ruledtabular}
\begin{tabular}{lccccc}
Uncertainty & CDF $\ell+$jets \cite{Aaltonen:2013vaf} & CDF $\ell\ell$ \cite{Aaltonen:2014eva} & D0 $\ell+$jets \cite{Abazov:2014oea} & D0 $\ell\ell$ \cite{Abazov:2013wxa} & Combination \\\hline
Statistical & 0.024 & 0.052 & 0.027 & 0.037 & 0.016 \\
Background & 0.015 & 0.029 & ${}^{+0.016}_{-0.018}$  & 0.008 & 0.008\\
Signal  & 0.007  & $<0.001$ & 0.008 & 0.005 & 0.006\\
Detector & 0.002 & 0.004 & ${}^{+0.008}_{-0.011}$ & 0.005  & 0.004\\
Method & ${}^{+0.013}_{-0.000}$ & 0.006 & 0.008 & 0.004 & 0.005\\
PDF & 0.003  & $<0.001$ & 0.002 &  $<0.001$ & 0.002\\
\end{tabular}
\end{ruledtabular}
\end{center}
\end{table*}

\begin{table*}[hbtp]
\begin{center}
\caption{Inputs to and results from the combination of the inclusive $\afblep$ asymmetries.}
\label{tab:AFBlepResults}
\begin{ruledtabular}
\begin{tabular}{lccccc}
\multirow{2}{*}{Analysis}& \multirow{2}{*}{$\afblep$} &\multicolumn{3}{c}{Uncertainty}&\multirow{2}{*}{Weight}\\\cline{3-5}
&& Stat. & Syst. & Total&\\\hline
CDF $\ell+$jets \cite{Aaltonen:2013vaf} & 0.105&0.024&${}^{+0.022}_{-0.017}$&${}^{+0.032}_{-0.029}$&0.40\\
CDF $\ell\ell$ \cite{Aaltonen:2014eva} & 0.072 & 0.052 & 0.030 & 0.060 & 0.11\\
D0 $\ell+$jets \cite{Abazov:2014oea} & 0.050 & 0.027 & ${}^{+0.020}_{-0.024}$ & ${}^{+0.034}_{-0.037}$ & 0.27\\
D0 $\ell\ell$ \cite{Abazov:2013wxa} & 0.044 & 0.037 & 0.011 & 0.039 & 0.23 \\ \hline
Combination & 0.073 & 0.016 & 0.012 & 0.020 &  \\
\end{tabular}
\end{ruledtabular}
\end{center}
\end{table*}

\begin{table*}[hbtp]
\begin{center}
\caption{Statistical and systematic uncertainties in the individual inclusive $\afbdeta$ inputs as well as in the combined results. 
}
\label{tab:AFBdetaUncertainties}
\begin{ruledtabular}
\begin{tabular}{lccc}
Uncertainty & CDF $\ell\ell$ \cite{Aaltonen:2014eva} &  D0 $\ell\ell$ \cite{Abazov:2013wxa} & Combination \\\hline
Statistical & 0.072 & 0.054 & 0.043\\
Background & 0.037 & 0.009 & 0.013\\
Signal & $<0.001$ & 0.009 & 0.001 \\
Detector & 0.003 & 0.006 & 0.008 \\
Method & 0.013 & 0.004 & 0.005 \\
PDF & $<0.001$ & $<0.001$ & 0.001 \\
\end{tabular}
\end{ruledtabular}
\end{center}
\end{table*}

\begin{table*}[hbtp]
\begin{center}
 \caption{Inputs to and results from the combination of the inclusive $\afbdeta$ asymmetries.}
\label{tab:AFBdetaResults}
\begin{ruledtabular}
\begin{tabular}{lccccc}
\multirow{2}{*}{Analysis}& \multirow{2}{*}{$\afbdeta$} &\multicolumn{3}{c}{Uncertainty}&\multirow{2}{*}{Weight}\\\cline{3-5}
&& Stat. & Syst. & Total&\\\hline
CDF $\ell\ell$ \cite{Aaltonen:2014eva} & 0.076 & 0.072 & 0.037 & 0.082 & 0.32\\
D0 $\ell\ell$ \cite{Abazov:2013wxa} & 0.123 & 0.054 & 0.015 & 0.056 & 0.68 \\\hline
Combination & 0.108 & 0.043 & 0.016 & 0.046 & \\

\end{tabular}
\end{ruledtabular}
\end{center}
\end{table*}


\begin{thebibliography}{99}
\bibitem{rapidity} The rapidity $y$ is defined as $y(\theta,\beta) = 1/2 \cdot \ln[(1+\beta\cdot\cos(\theta)) / (1-\beta\cdot\cos(\theta))]$, where $\theta$ is the polar angle relative to the proton beam and $\beta$ is the ratio of a particle's momentum to its energy. The pseudo-rapidity is defined as $\eta = y(\theta,1)$.
\end{thebibliography}
\end{document}